\documentclass[a4paper]{jpconf}
\usepackage{graphicx}
\usepackage{icsghead}
\usepackage{bm}
\begin{document}
\title{Statistical inference of co-movements of stocks during a financial crisis}

\author{Takero Ibuki\dag, Shunsuke Higano\ddag, Sei Suzuki\S, Jun-ichi Inoue\pounds \,\,and\,\,
Anirban Chakraborti\P}

\address{\dag\,Data Mining Group, 
Service $\&$ Solution Development Department,  
Research and Development Center, NTT DOCOMO, INC.,  
3-6 Hikarino-oka, Yokosuka-shi, Kanagawa 239-8536, Japan \\
\ddag\,
Hokkaido  
Prefectural Police, 
N2-W7, Chuo-Ku, Sapporo 060-8520, Japan \\
\S\,Department of Basic Sciences, Saitama Medical University, 
38 Morohongo, Moroyama, Saitama 350-0495, Japan \\
\pounds\,Graduate School of Information Science and Technology, 
Hokkaido University, N14-W9, Kita-Ku, Sapporo 060-0814, Japan \\
\P\,Laboratoire de
  Math\'{e}matiques Appliqu\'{e}es aux Syst\`{e}mes, \'{E}cole
  Centrale Paris, 92290 Ch\^{a}tenay-Malabry, France
}

\ead{\dag\,takerou.ibuki.xt@nttdocomo.com,
\ddag\,higano@complex.ist.hokudai.ac.jp, 
\pounds\,j$\underline{\,\,\,}$inoue@ist.hokudai.ac.jp, jinoue@cb4.so-net.ne.jp, 
\S\,sei01@saitama-med.ac.jp, 
\P\,anirban.chakraborti@ecp.fr}

\begin{abstract}
In  order to figure out and to forecast the 
emergence phenomena of social systems,  
we propose several probabilistic models for the analysis of financial markets, especially 
around a crisis. 
We first attempt to visualize the collective behaviour of markets during a financial crisis 
through cross-correlations between typical Japanese daily stocks 
by making use of multi-dimensional scaling. 
We find that all the two-dimensional points (stocks) 
shrink into a single small region when a economic crisis takes place. 
By using the properties of cross-correlations in financial markets especially during a crisis, 
we next propose a theoretical framework to predict several time-series 
simultaneously. 
Our model system is basically described by a variant of the multi-layered Ising model 
with random fields as non-stationary time series. 
Hyper-parameters appearing in the probabilistic model 
are estimated by means of minimizing the `cumulative error' in the past market history. 
The justification and validity of our approaches are numerically examined 
for several empirical data sets. 
\end{abstract}

\section{Introduction}
\label{sec:1}
One of central modern issues in quantitative finance 
is to determine to what extent the market is `efficient'; 
crudely, 
whether 
there is an equal chance 
that the stock is under or over value at any time point. 
From the view point of statistics, 
the market is regarded as efficient 
when the market price is an unbiased estimate, 
in other words, 
when the price can be greater or less than the true 
value as long as the deviation is completely random. 

Recently, as huge high-frequency financial data sets can be stored and analysed, 
the so-called `stylized (empirical) facts' \cite{Cont,AC1,AC2,AC3}
such as heavy tails of returns, 
volatility clustering, gain/loss asymmetry {\it etc.}  have been found, 
in particular, in the research field of econophysics \cite{Bouchaud,Stanley,Enrico,Aoyama,Sinha,CUP2013,Wiley2006}. 
At the same time, 
several empirical facts provide an evidence 
to show that 
there exist some `seasonal effects' in financial market (the so-called `market anomaly'). 
In fact, 
it is well-known that 
buying stocks at the end of year and 
selling them at the beginning of the next year 
is sometimes 
less risky (the so-called {\it January effects} \cite{Keim}). 
Hence, 
it is now partially accepted that the financial market 
is `weakly' efficient or it is sometimes 
inefficient 
to some extent and in certain time scale. 

Actually, there are several evidences 
to show the market inefficiency during financial crisis. 
At financial crisis, traders are more likely to behave according to the `mood' (atmosphere) in society (financial market) and 
they incline to take rather `irrational' strategies in some sense. 
Thus, this collective behavior might cause some market anomaly. 

In the literature of behavioral economics \cite{Kahneman}, a concept of the 
so-called {\it information cascade} (or {\it Herding effect}) is well-known as a result of 
such human (traders') collective behaviour. One of the key measurements  
to understand such financial cascade 
is `correlation' between ingredients 
in the societies. 
For instance, cross-correlations between stocks, 
traders are quite important to figure out the human collective phenomena. 
As the correlation could be found in various scale-lengths, 
from macroscopic stock price level to microscopic trader's level,   
the cascade also might be 
observed `hierarchically' in such various scales 
from prices of several stocks to  
ways (strategies) of trader's decision making. 
Actually, we sometimes encounter the problem to find unusual structure in correlated time series observed from multi-dimensional information channels. 
Such time series obtained from multi-channel measurement have been widely provided in both natural and social sciences. 
Hence, it is now quite important for us to carry out empirical data analysis extensively to solve various modern and serious problems around us.

After the earthquake on 11th March 2011, Japanese NIKKEI stock market quickly responded to the crisis and quite a lot of traders 
sold their stocks of companies whose branches or plants are located in that disaster stricken area. 
As the result, the Nikkei stock average suddenly drops after the crisis \cite{ISI}. 
\begin{figure}[ht]
\begin{center}
\includegraphics[width=8.1cm]{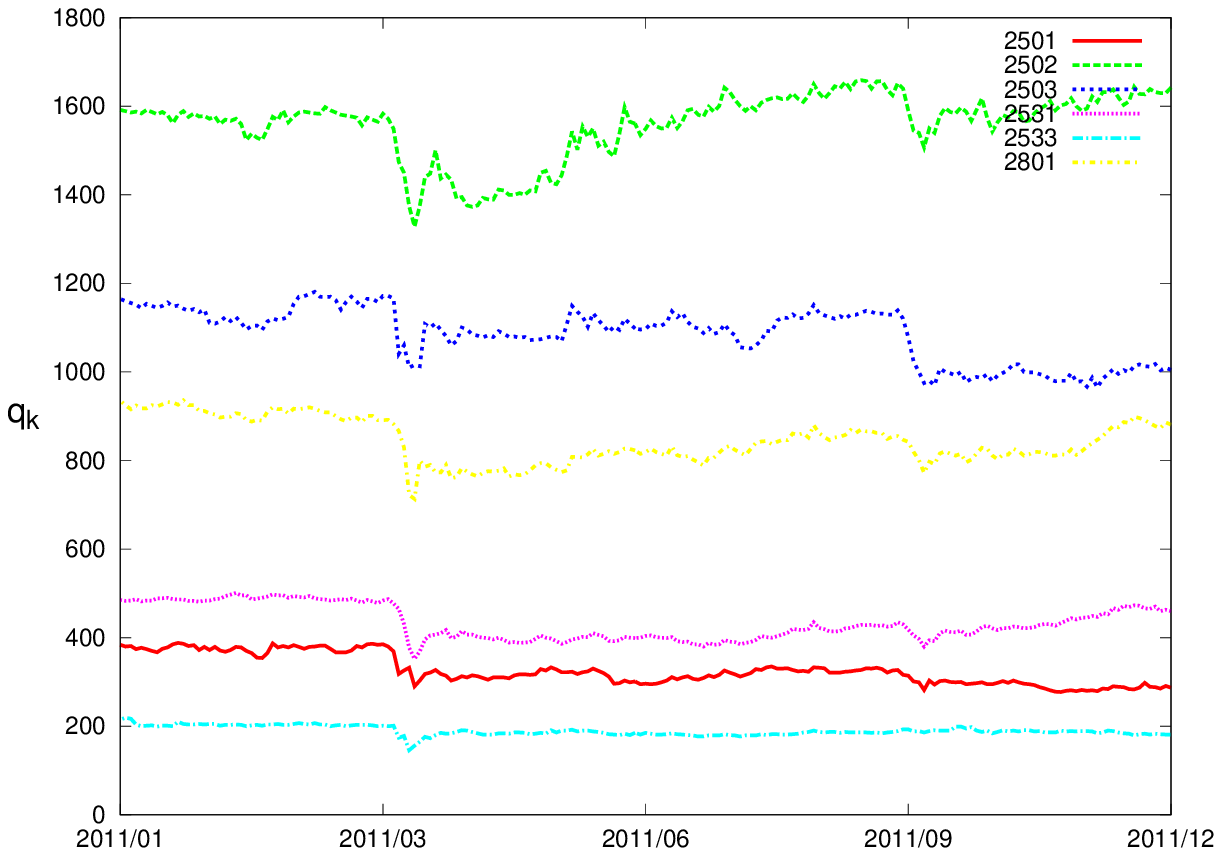} \hspace{-0.5cm}
\includegraphics[width=8.1cm]{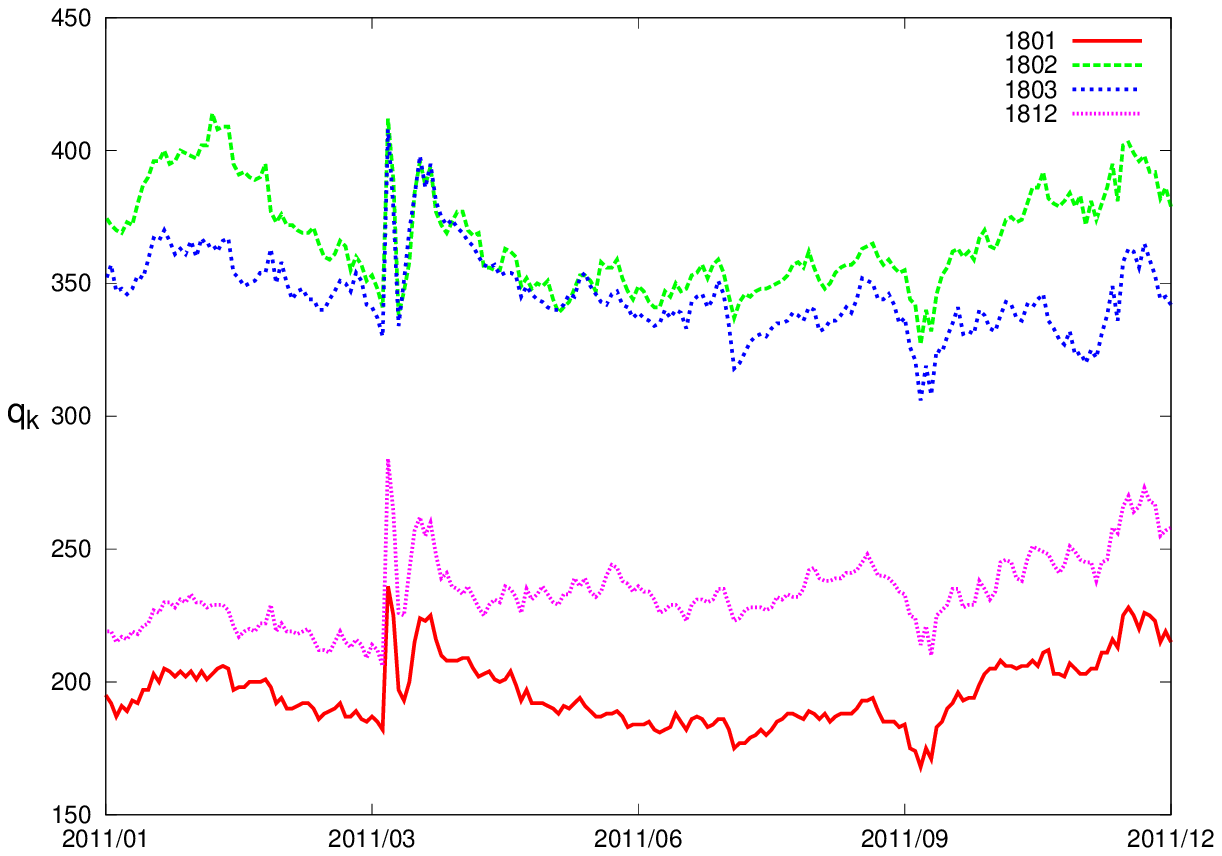}
\end{center}
\caption{\footnotesize 
The prices of several major stocks in food industries (left) and 
construction industries  (right) as a function of time (these are daily data sets). 
Each number shown at each line caption denotes the ID for each company: 
({\it i.e.} {\tt 2501}: Sapporo Breweries, 
{\tt 2502}: Asahi Breweries, {\tt 2503}: Kirin Holdings, 
{\tt 2531}: Takara Holdings, {\tt 2533}: Oenon Holdings, 
{\tt 2801}: Kikkoman Corporation, 
{\tt 1801}: Taisei Corporation, 
{\tt 1802}: Obayashi Corporation, 
{\tt 1803}: Shimizu Corporation, 
{\tt 1812}: Kajima Corporation. 
These IDs can be checked at the web site \cite{Yahoo}) (COLOR ONLINE)
}
\label{fig:fg001}
\end{figure}
However, it is impossible for us to mention the co-movement of stocks having correlation (with the majority bulk component including themselves) or anti-correlation 
during the crisis. 
In Figure \ref{fig:fg001}, 
we plot the prices of several major stocks in food business (left) and 
construction business (right) as a function of time (these are daily data sets).
From this figure, we find that 
The price of stocks for the same type of business behaves 
as correlated time series, 
whereas for different types of business, 
say, food industries and construction business, 
they have apparent any-correlations 
especially during the crisis.

Hence, it might be quite important for us to make an attempt to bring out more `microscopic' useful information, 
which is never obtained from the averaged macroscopic quantities such as stock average, 
about the market.  As a candidate of such `microscopic information', we can use the (linear) correlation coefficient 
based on the two-body interactions between stocks \cite{ISI,Mantegna,Anirban,Anirban1,Anirban2,Anirban3,Anirban4,Ibuki}. 
To make out the mechanism of financial crisis, it might be helpful for us to visualize 
such correlations in stocks and compare the dynamical behaviour of the correlation before and after crisis.

In this paper, in order to show and explain 
the hierarchical information cascade, we visualize the correlation of each stock in two-dimension. 
We specify each location of $K$ stocks from a given set of the 
$K(K-1)/2$ distances by making use of the so-called multi-dimensional scaling (MDS) \cite{MDS}. 
We also propose a theoretical framework to predict several time-series 
simultaneously by using cross-correlations in financial markets. 
The justification of this assumption is numerically checked for the empirical Japanese stock data, 
for instance, those around 11 March 2011, 
and for foreign currency exchange rates around Greek crisis in spring 2010. 

This paper is organized as follows. 
In section \ref{sec:se2}, 
we explain our tools of analysis, namely, the correlation coefficient and 
multi-dimensional scaling. 
In section \ref{sec:se4}, 
our forecasting model for stock prices is introduced. 
In section \ref{sec:se5}, 
we examine our model for empirical data sets. 
The last section gives several remarks. 
\section{Linear correlation coefficient and multi-dimensional scaling}
\label{sec:se2}
We utilize the linear correlation coefficient 
to measure the strength of correlation between stocks \cite{Mantegna,Anirban,Anirban1,Anirban2,Anirban3,Anirban4}. 
The correlation coefficient (Pearson estimator) is calculated as follows. 

Let us define $p_{t}^{(i)} (\geq 0)$ as a price of stock $i$ at time $t$. 
Then, we evaluate the return of the price $p_{t}^{(i)}$ in terms of the logarithmic measurement as 
\begin{equation}
\Delta r_{i} (t) \equiv  
\log p_{t}^{(i)}-\log p_{t-1}^{(i)}.
\end{equation} 
For the above logarithmically rescaled return, we calculate the moving average over a time window (interval) with width $M$ as
\begin{equation}
\overline{\Delta r_{i} (t)}
 \equiv  
 \frac{1}{M}
\sum_{l=t-M+1}^{t} 
\Delta r_{i} (l) 
\end{equation}
for stock $i$, 
and we also evaluate the two-body correlation between 
stocks $i,j$ by the following definition 
\begin{equation}
\overline{\Delta r_{i}(t)
\Delta r_{j} (t)}
 \equiv 
\frac{1}{M}
\sum_{l=t-M+1}^{t}
\Delta r_{i} (l)
\Delta r_{j} (l).
\end{equation}
Then, the linear correlation coefficient 
is given by
\begin{eqnarray}
c_{ij} (t) & = & 
\frac{
\overline{\Delta r_{i} (t)
\Delta r_{j} (t)}
 - 
(\overline{\Delta r_{i} (t)})
(\overline{\Delta r_{j} (t)})
}
{
\sqrt{
[
\overline{(\Delta r_{i} (t))^{2}}
-(\overline{\Delta r_{i} (t)})^{2}
]
[
\overline{
(\Delta r_{j} (t))^{2}}
-(\overline{
\Delta r_{j} (t)})^{2}]
}}
\label{eq:coef}. 
\end{eqnarray}
\subsection{Distribution of linear coefficient}
As empirical data set, we pick up 200 stocks including 
the so-called TOPIX (\underline{TO}kyo stock \underline{P}rice \underline{I}nde\underline{X}) Core30, 
which consists of typical 30 stock indices being picked up from the view point 
of `current price' or `liquidity'  
from the Japanese Nikkei stock market \cite{Yahoo}. 
It should be kept in mind that 
the data itself is not provided as `tick-by-tick' intra-day data, 
the minimal time interval of the data is one day 
(the closing price is given in the data set). 

\begin{figure}[ht]
\begin{center}
\includegraphics[width=12cm]{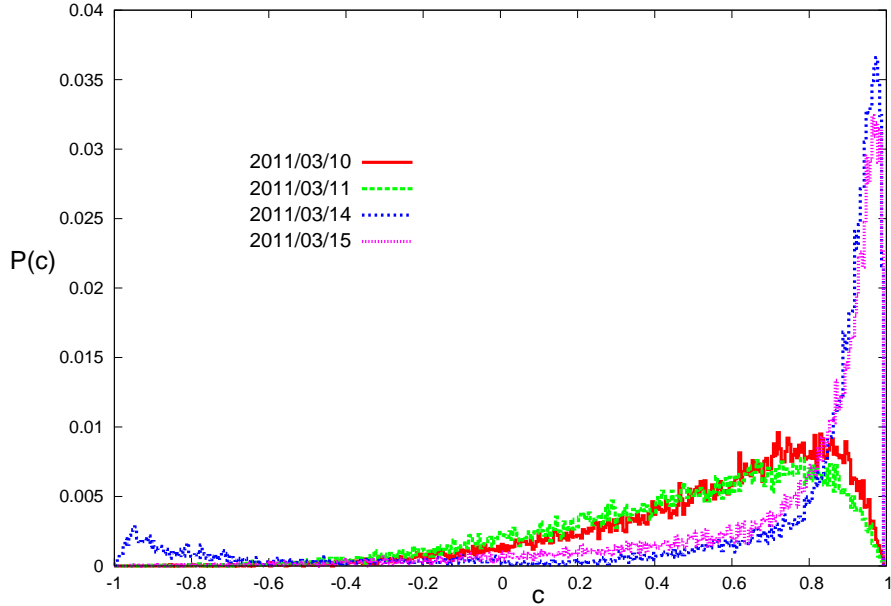}  
\end{center}
\caption{\footnotesize
The distribution $P(c)$ 
of linear correlation coefficients. 
The result for before crisis (10th, 11th March 2011) and 
after crisis (14th, 15th March 2011) are shown. 
It should be noted that 
12th and 13th were weekend and 
the market was closed. 
We find that a single peak before the crisis 
splits into the `correlated' (with the majority bulk component including themselves) and `anti-correlated' clusters on 14th. 
However, the anti-correlated cluster disappears on the next day. 
(COLOR ONLINE)
}
\label{fig:fg2b}
\end{figure}

In order to investigate the statistical properties, 
we evaluate the distribution of 
the correlation coefficient $P(c)$ and 
plot the result in Figure \ref{fig:fg2b}.
From this figure, we clearly find that 
the distribution is skewed and possesses a single peak at 
$c >0$  before the crisis. 
Namely, most of the 200 stocks are mutually correlated. 
On the other hand, just after the crisis, 
say, 14th March 2012, the single peak splits into 
two components and the bulk in which 
some pairs of two stocks posses the negative correlation appears. 
Thus, one can grasp the macroscopic aspect of the collective behaviour 
of the 200 stocks for both before and after the crisis. 
However, much more microscopic properties 
of the 200 stocks are unfortunately hided behind the distribution $P(c)$. 

To reveal such hidden microscopic aspects of the collective behaviour 
of the 200 stocks, we shall next attempt to visualize the 
relationship between these stocks by specifying the location 
of each stock in two-dimensional space. 
\subsection{The multi-dimensional scaling: From correlation to distance}
To make a plot of the location of 
each stock, one needs the information about 
the Euclidean distance between arbitrary two stocks. 
As we saw in the previous subsection, 
the correlation coefficient $c_{ij} (t)$ might posses 
some useful information about the relationship 
between arbitrary two stocks $i,j$, however, 
it should be noticed that 
the correlation coefficient (\ref{eq:coef}) 
satisfies $-1 \leq c_{ij} (t) \leq 1$, 
and apparently it cannot be treated as a `distance'. 
Hence, here we transform 
the correlation coefficient $c_{ij} (t)$ into the distance $d_{ij} (t)$ between 
the stocks 
$i,j$ as 
\begin{equation}
d_{ij} (t)  =  
\sqrt{\frac{1-c_{ij} (t)}{2}}.
\end{equation}
We should bear in mind that 
the above distance satisfies 
$0 \leq d_{ij} (t) \leq 1$ and 
defines the metric space in multi-dimension \cite{ISI}. 

Obviously, once we obtain the location 
vectors $\bm{X}_{i}, \bm{X}_{j}$ for the stocks $i,j$ , 
one can easily calculate the 
distance between them 
as $\|\bm{X}_{i}-\bm{X}_{j}\|$. 
However, the inverse process, 
namely, to specify the 
location vectors $\bm{X}_{i}, \bm{X}_{j}$ 
for a given distance $\|\bm{X}_{i}-\bm{X}_{j}\|$ 
is not so easy when the number of the stocks $N$ increases. 
To carry out the inverse process systematically, 
we can use the well-known 
method named as 
{\it multi-dimensional scaling (MDS)} (see {\it e.g.}  \cite{MDS}). 
In following, we explain the procedure. 

Let us first specify the location of 
an arbitrary stock $i$ at time $t$ by means of a $P$-dimensional vector $\bm{X}_{i}^{(t)}$ 
in the following way (Hereafter, we consider the case of $P=2$ especially): 
\begin{equation} 
\bm{X}_{i} (t) \equiv (x_{i1} (t),x_{i2} (t),\cdots,x_{iP} (t)),\,\,i=1,\cdots,K.
\end{equation}
Naturally, 
the Euclidean distance 
between 
arbitrary two stocks $i,j$ is 
now given by 
\begin{equation}
d_{ij} (t) = 
\sqrt{\sum_{m=1}^{P}
(x_{im} (t)-x_{jm} (t))^{2}}.
\end{equation} 
Hence, the inner product of 
location vectors of stocks $i$ and $j$ is 
also calculated as  
\begin{equation}
2\{\bm{X}_{i} (t) \cdot 
\bm{X}_{j} (t)\} = 
2z_{ij} (t) = 
\frac{1}{K}
\sum_{i=1}^{K}
d_{ij} (t)^{2}+
\frac{1}{K}
\sum_{j=1}^{K}
d_{ij} (t)^{2} - 
\frac{1}{K^{2}}
\sum_{i,j=1}^{K}
d_{ij} (t)^{2} -d_{ij} (t)^{2}
\end{equation}
where we should notice that 
we chose the origin of 
axis as the `center of mass' 
$\bm{X}_{c} (t)$ 
for 
$K$ stocks points, 
that is to say, 
\begin{equation}
\bm{X}_{c} (t) \equiv 
\frac{1}{K} 
\sum_{i=1}^{K}\bm{X}_{i} (t) = 
\bm{0}
\label{eq:origin}
\end{equation}
to specify an arbitrary stock (vector) at each time $t$. 
This equation (\ref{eq:origin}) implies that 
the center of mass is a time-independent vector and it is definitely fixed at the origin for all time $t \geq 0$.

Then, in order to look for 
the locations  
$\bm{X}_{i} (t),\, i=1\,\cdots,K$ which generates 
a set of distances $\{d_{ij} (t)\}$ consistently, we 
should minimize 
the following 
energy function (cost): 
\begin{equation}
E_{t} = 
\sum_{i,j=1}^{K}
\left(
z_{ij} (t)-\sum_{m=1}^{P}x_{im} (t)
x_{jm} (t)
\right)^{2}
\end{equation}
with respect to $\bm{X}_{i} (t), \,i=1,\cdots,K$. 

\begin{figure}[ht]
\begin{center}
\includegraphics[width=7.8cm]{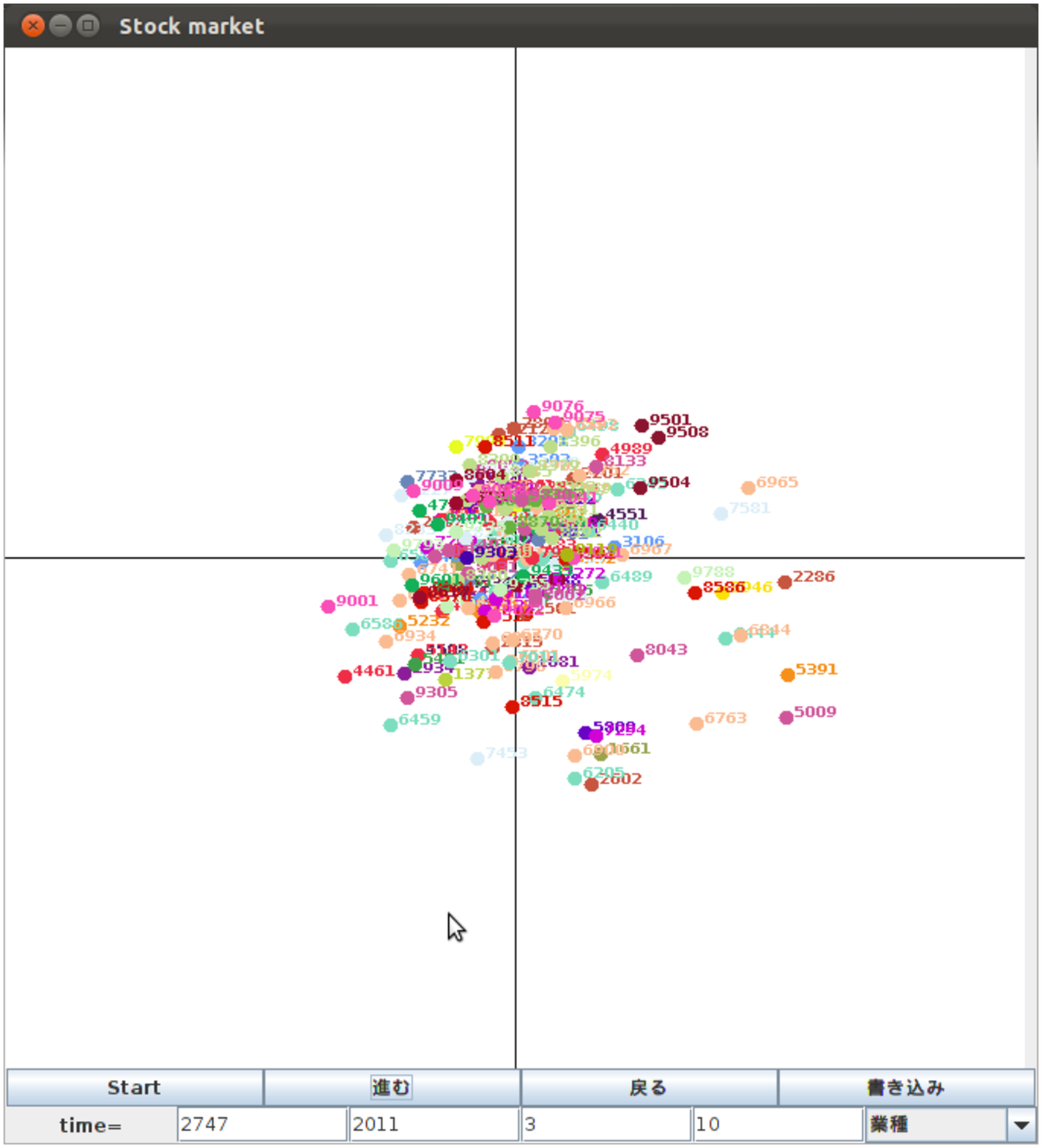}  
\mbox{}\hspace{-0.1cm}\includegraphics[width=7.8cm]{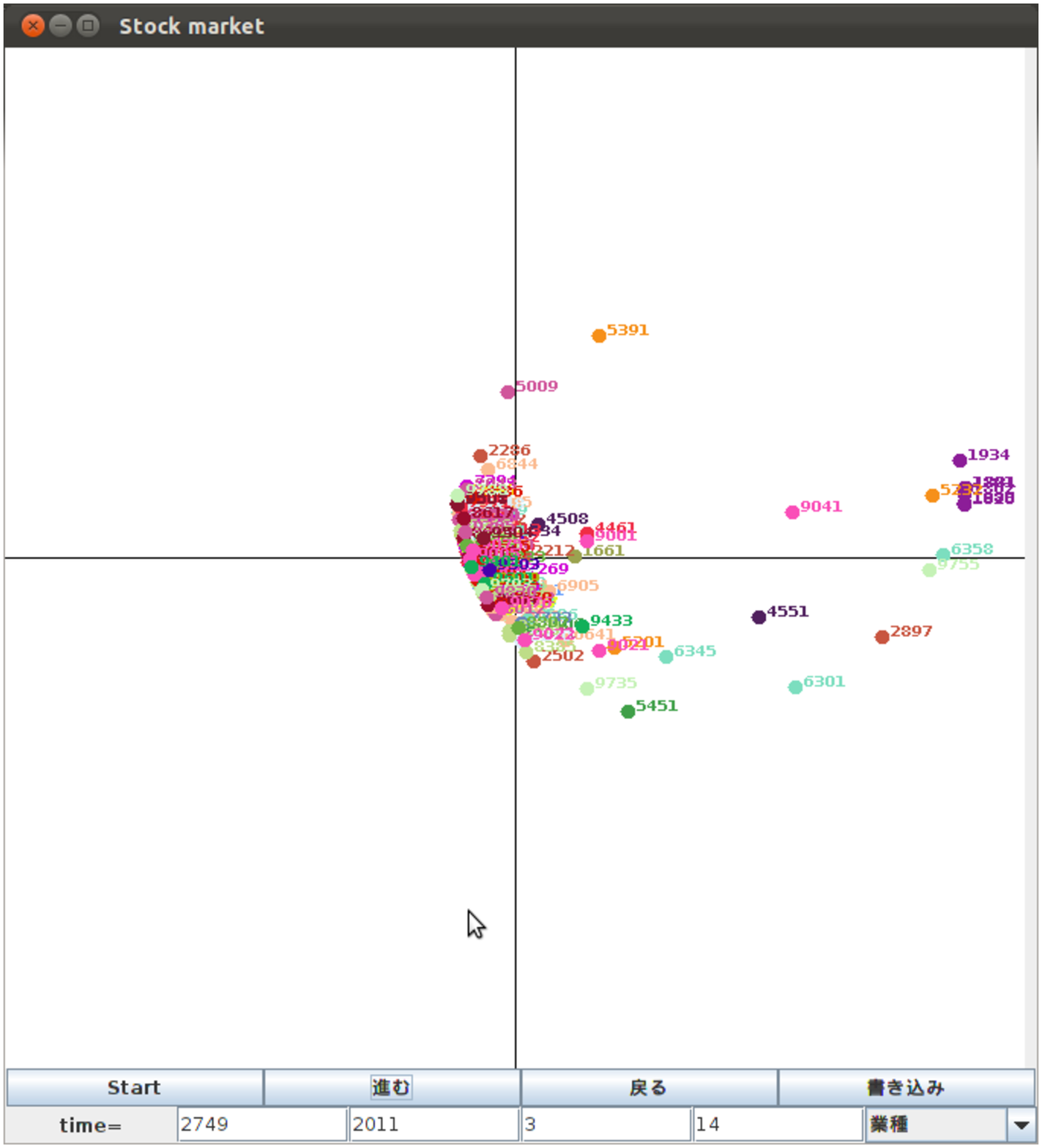} 
\end{center}
\caption{\footnotesize
The result of the MDS. We pick up 200 stocks including 
the so-called TOPIX Core30 and 
the Nikkei stock average as 
empirical data set. 
The results on 
10th (left) and 14th (right) March 2011 
are shown. 
Different colors indicate different types of
business. The numbers accompanying the dots
show company IDs. 
The isolated points anti-correlated with the bulk are 
construction business, 
such as 
{\tt 1934}: Yurtec Corporation, 
which is a construction industry in Tohoku area, 
{\tt 1826}: 
Sata Construction Co. Ltd. 
We set the width of time window 
to evaluate the correlation coefficient 
as $M=7$ (days). (COLOR ONLINE)}
\label{fig:fg2}
\end{figure}

Thus, our problem to find the best possible locations for 
stocks is now rewritten in terms of 
an optimization problem to look for the solution $\bm{X}_{i} (t), \,i=1,\cdots,K$ that minimizes  
the energy function $E_{t}$ at each time step $t$. 
We plot the result in Figure \ref{fig:fg2} for 
the same data set used in the plot of 
Figure \ref{fig:fg2b}. 
From these panels, we clearly find that 
after the crisis, 
the scattered plots actually shrink into a small limited region 
centered at the origin (the center of mass) as we expected before.  
Isolated several dots separating from main clusters 
denote the price of `building industry'. 
These isolated points anti-correlated with the bulk are 
construction business, 
such as 
ID {\tt 1934}: Yurtec corporation, 
which is a construction industry in Tohoku area, 
ID {\tt 1826}: 
Sata construction Co. Ltd. 
Apparently we recognize that the large amount of 
demand is expected in such industries even after the crisis (mega earthquakes).  
\section{Prediction by using cross-correlation}
\label{sec:se4}
It is necessary for us to 
investigate the human collective behaviour of the 
social agents such as traders 
in order to construct persistent 
systems in earthquake disasters. 
In financial systems, 
the price of each commodity as a 
macroscopic quantity 
is determined by huge number of trader's making decisions. 
To predict the price efficiently, 
a lot of mathematical tools 
such as AR model and its extensions called as 
ARCH model or GARCH model \cite{Stanley}, 
or Kalman filter and its various variants have been proposed. 
Recently, 
besides these rather traditional models, 
several physics inspired models have been also 
introduced by several authors \cite{Bouchaud2,Masukawa,PUCK}. 
However, 
these models apparently lack the microscopic view point. 
Namely, 
in these models, the predicted price is not constructed by 
the result from microscopic trader's decision making. 

Among these studies of prediction of 
stock prices, 
Kaizoji \cite{Kaizoji2000} 
attempted to 
represent `buying' and `selling' signals 
posted by traders 
by `Ising spin'. 
He proposed a model in which 
the return of the price 
is determined by the `magnetization' of the Ising system. 
He carried out computer simulations 
and concluded that 
there are interesting relationship 
between financial phenomena 
such as `bubble' or `crush' and 
physically collective phenomena which 
are referred to as `phase transition' in the Ising magnetic system. 

However, those studies seem to be not yet extensive and 
there exist several open questions to be clarified. 
For instance, 
dynamics of 
macroscopic quantities which might specify 
the collective behaviour of traders 
should be revealed more extensively. 
And as we saw in the previous sections, 
just after crisis, 
several stocks are strongly correlated. 
Hence, we might use the cross-correlation 
to predict several prices simultaneously 
by modifying the Kaizoji's model \cite{Kaizoji2000} 
by means of the cross-correlations. 
Therefore, 
from this section, we are focusing on 
the prediction of 
prices of several stocks simultaneously 
by using the cross-correlation in stock markets.  
\subsection{A link from microscopic making decisions to macroscopic prices}
In order to investigate the effect of the cross-correlation 
on the prediction of several stock-prices, 
we first explain the model proposed by Kaizoji \cite{Kaizoji2000} as a basic model. 

Let us define $p_{t}^{(k)}$ 
as the price of commodity $k$ at time $t$. 
Then, the return, which is 
defined as the difference between prices at successive two time steps 
$t$ and $t+1$ is given by 
\begin{equation}
\Delta p_{t}^{(k)} \equiv p_{t+1}^{(k)}-p_{t}^{(k)}=m^{(k)}_{t},\,\,\,
k=1,\cdots, K. 
\label{eq:dyn}
\end{equation}
In order to construct the return $m_{t}^{(k)}$ from 
the microscopic view point, 
we assume that 
each trader ($i=1,\cdots,N_{k}$) 
possessing the stock $k$  buys or sells 
$v_{it}^{(k)}$-volumes at each time step $t$. 
Then, let us call the group of buyers as 
$\mathcal{A}_{+} (k,t)$, whereas 
the group of sellers is referred to as 
$\mathcal{A}_{-} (k,t)$. 
Hence, the total volumes of buying and selling are explicitly given by 
\begin{equation}
\psi_{+} (k,t) \equiv   \sum_{i \in \mathcal{A}_{+} (k,t)}v_{it}^{(k)},\,\,
\psi_{-} (k,t)  \equiv   \sum_{i \in \mathcal{A}_{-} (k,t)}v_{it}^{(k)}, 
\label{eq:psi}
\end{equation}
respectively. 
We should keep in mind that 
the total number of traders dealing with the stock $k$ 
is conserved, namely, the condition: 
\begin{equation}
\mathcal{A}_{+} (k,t) + 
\mathcal{A}_{-} (k,t) = N_{k}
\end{equation}
should be satisfied. 

Then, the return $m_{t}^{(k)}$ 
is naturally defined by means of (\ref{eq:psi}) 
\begin{equation}
m_{t} = \lambda (\psi_{+} (k,t)-\psi_{-} (k,t))
\end{equation}
where $\lambda$ is a positive constant. 
Namely, 
when the volume of 
buyers 
is greater than that of sellers, 
$\psi_{+} (k,t) > \psi_{-} (k,t)$, 
the return becomes positive $m_{t}^{(k)} >0$. 
As the result, the price of the commodity $k$ 
should be increased at the next time step 
as $\Delta p_{t}^{(k)} = m_{t}^{(k)}$. 
\subsubsection{Ising spin representation}
The above microscopic observation and the set-up might be naturally accepted, 
however, here we shall 
make the situation much more simpler. Namely, 
we omit the information about the volume by setting  
$v_{it}^{(k)}=1\,(\forall_{i,t,k})$. 
The making decision of each trader ($i=1,\cdots,N_{k}$) is now 
obtained simply by an Ising spin:  
\begin{equation}
S_{i}^{(k)}(t) =
\left\{ 
\begin{array}{cc}
+1  & \mbox{(the trader $i$ buys the stock $k$ at time $t$)} \\
 -1 &  \mbox{(the trader $i$ sells the stock $k$ at time $t$)}
 \end{array}
 \right.
 \end{equation} 
The return is also simplified as  
\begin{equation}
m_{t}^{(k)} =\lambda (\psi_{+} (k,t)-\psi_{-} (k,t)) = 
\lambda
\sum_{i=1}^{N_{k}}v_{it}^{(k)} S_{i}^{(k)}(t)
\end{equation}
where we set 
$\lambda=N_{k}^{-1}$ to 
make the return: 
\begin{equation}
m_{t}^{(k)} = 
\frac{1}{N_{k}}
\sum_{i=1}^{N}
S_{i}^{(k)}(t)
\label{eq:update_pt}
\end{equation}
satisfying $|m_{t}^{(t)}|\leq 1$. 
Thus, $m_{t}^{k}$ corresponds to the `magnetization' 
in statistical physics, and 
the update rule of the price for the stock $k$ 
is governed in terms of the magnetization $m_{t}^{(k)}$ as 
\begin{equation}
\Delta p_{t}^{(k)} = m_{t}^{(k)},\,k=1,\cdots,K. 
\end{equation}
\subsection{The energy function}
We next introduce the energy function of the system.
\begin{equation}
E_{t}(\bm{S}^{(k)})  =     -\frac{J_{t}^{(k)}}{N_{k}}\sum_{ij}S_{i}^{(k)}S_{j}^{(k)}  -  
h_{t}^{(k)}
\sum_{i}
\sigma_{\tau}^{(k)}(t) S_{i}^{(k)}  -    \gamma_{t}^{(k)}\sum_{i}
\left(
\frac{1}{K}\sum_{\mu \neq k}^{K}
c_{k\mu} (t) m_{t}^{(\mu)}
\right)S_{i}^{(k)}
\label{eq:energy2_new}
\end{equation}
where 
we omitted the time dependence in $S_{i}^{(k)}$ for simplicity. 
The first term in the right hand side of (\ref{eq:energy2_new}) induces human collective behaviour, 
namely, each agent inclines to take the same decision as the others 
to decrease the total energy. 
The effect of this first term on the minimization of 
total energy might be recognized as 
the so-called {\it Keynes's beauty contest}. 
It means that traders tend to make the same decision 
as the others, in particular, during a crisis. 
Namely, when a big negative news (mood) such as 
earthquake in Japan 
is broadcasted, 
almost all of the traders might sell 
their own stocks as the others also sell without any rationality. 
In other words, 
the first term causes 
the human collective behaviour of the traders, 
which is sometimes refereed to as information cascade in the literature of 
behavioral economics, for 
$J_{t}^{(k)} \geq 
h_{t}^{(k)}, \gamma_{t}$. 
Actually, we find that 
the lower bound of the first term in the right hand side of 
(\ref{eq:energy2_new}) is evaluated as  
\begin{equation}
-\frac{J_{t}^{(k)}}{N_{k}}
\sum_{ij}S_{i}^{(k)}
S_{j}^{(k)} \geq 
-N_{k} J_{t}^{(k)}
\end{equation}
where 
the equal sign is satisfied 
if and only if 
$S_{i}^{(k)}=1\,\,
(\forall_{i})$ or 
$S_{i}^{(k)}=-1\,\,
(\forall_{i})$ holds. 

The second term in (\ref{eq:energy2_new}) 
represents the 
cross-correlation between the decisions of traders  
and market (historical) information. 
In this paper, we choose the `trends' : 
\begin{equation} 
\sigma_{\tau}^{(k)}(t)  \equiv \frac{p_{t}-p_{t-\tau}}{\tau}
\label{eq:trend}
\end{equation}
for such information. 
This means that 
the total energy should decrease 
when each trader 
posts the sign 
$S_{i}^{(k)}=\pm 1$ 
to the market so as to 
make the product 
$\sigma_{\tau}^{(k)}(t) S_{i}^{(k)}$ definitely positive. 
In other words, 
each trader buys the commodity $k$ if the price tends to 
increase over the past $\tau$-time steps, 
whereas 
the trader sells the stock vice versa.  
Actually, 
the second term is definitely minimized as 
\begin{equation}
-h_{t}^{(k)} \sum_{i}
\sigma_{\tau}^{(k)}(t) S_{i}^{(k)} \geq 
-N_{k} h_{t}^{(k)} 
|\sigma_{\tau}^{(k)}(t)|
\end{equation}
where the equality should be satisfied when 
${\rm sgn}[S_{i}^{(k)}] = 
{\rm sgn}[\sigma_{\tau}^{(k)}(t)]$ holds. 

The third term, which does not appear in the references \cite{Kaizoji2000}, comes from correlation 
between the trader $i$ possessing the commodity $k$ and 
 a `typical trader' (a mean-field) $m_{t}^{(\mu)}$ possessing stocks $\mu\, (\neq k)$. A factor $c_{k \mu} (t)$ appearing in 
 the third term 
 stands for the correlation 
 coefficient and we should remember that it is given by (\ref{eq:coef}) explicitly. 
 Hence, 
 when the stocks 
 $k$ and $\mu$ are correlated 
 in terms of the positive coefficient, 
 namely, $c_{k \mu} (t)>0$ 
 from time $t-M+1$ to time $t$, 
 the trader possessing 
 the stock $k$ inclines to take 
 the same decision as the typical trader dealing with the stock $\mu$, that is, $m_{t}^{(\mu)}$. 
 
 From the above argument, 
 it might be useful for us to investigate to 
 what extent those traders behave collectively 
 through the values of 
 macroscopic hyper-parameters 
 $J_{t}^{(k)}, h_{t}^{(k)}$ and 
 $\gamma_{t}^{(k)}$. 
 Namely, in the realistic stock market, there might exist a possibility that 
 human behaviour turns out to 
 be collective and `irrational' in some sense 
 when these parameters change 
 the values as $h_{t}^{(k)}/J_{t}^{(k)} \to 0, \gamma_{t}^{(k)}/J_{t}^{(k)} \to 0$ and 
 $J_{t}^{(k)} \to 1$ (a critical point of infinite range ferromagnetic Ising model). 
\subsection{The Boltzmann-Gibbs distribution}
It should be noticed that 
the state vectors of the agents: $\bm{S}^{(k)}=(S_{1},\cdots,S_{N_{k}}),\,\,
k=1,\cdots,K$ 
are determined so as to minimize 
the energy function (\ref{eq:energy2_new}) 
from the argument in the previous subsection. 
For most of the cases, 
the solution should be unique. 
However, 
in realistic financial markets, 
the decisions by agents should be much more `diverse'. 
Thus, here we consider 
statistical ensemble of traders for each commodity
$\bm{S}^{(k)}$ 
and define the distribution of 
the ensemble by $P(\bm{S}^{(k)})$. 
Then, we shall look for the 
suitable distribution 
which maximizes the so-called Shannon's entropy
\begin{equation}
H=-\sum_{\bm{S}^{(k)}}P(\bm{S}^{(k)}) \log P(\bm{S}^{(k)})
\end{equation}
under two distinct constraints: 
\begin{equation}
\sum_{\bm{S}^{(k)}}P(\bm{S}^{(k)})=1,\,\,
\sum_{\bm{S}^{(k)}}P(\bm{S}^{(k)})E(\bm{S}^{(k)})=E_{k}.
\end{equation}
Namely, according to Jaynes \cite{Jaynes}, we choose the distribution 
which minimizes the following functional $f_{k} \{P(\bm{S}^{(k)})\}$: 
\begin{eqnarray}
f_{k} \{P(\bm{S}^{(k)})\} & = &  
-\sum_{\bm{S}^{(k)}}
P(\bm{S}^{(k)}) 
\log P(\bm{S}^{(k)})  - 
\lambda_{1}^{(k)}
\left(
\sum_{\bm{S}^{(k)}}
P(\bm{S}^{(k)}) -1
\right) \nonumber \\
\mbox{} & - &   
\lambda_{2}^{(k)} 
\left(
\sum_{\bm{S}^{(k)}}
P(\bm{S}^{(k)}) 
E_{k}(\bm{S}^{(k)})-E_{k}
\right)
\end{eqnarray}
where $\lambda_{1}^{(k)}, 
\lambda_{2}^{(k)}$ are 
Lagrange's multipliers. 
After some easy algebra, we immediately notice 
that $\lambda_{1}^{(k)}$ gives a normalization constant 
of $P(\bm{S}^{(k)})$ and 
$\lambda_{2}^{(k)} \equiv \beta$ is a control parameter for `thermal fluctuation' in the system. 
Thus, we have 
the solution as 
\begin{equation}
P(\bm{S}^{(k)}) = 
\frac{{\exp}[-\beta E(\bm{S}^{(k)})]}
{\sum_{\bm{S}^{(k)}} {\exp}[-\beta E(\bm{S}^{(k)})]}
\label{eq:gibbs}
\end{equation}
where 
$\beta$ stands for the inverse-temperature, 
and hereafter we set $\beta=1$. 
We also defined 
the sum with respect to 
$\bm{S}^{(k)}$ by 
\begin{equation}
\sum_{\bm{S}^{(k)}}(\cdots) \equiv 
\sum_{S_{1}^{(k)}=\pm1}\cdots 
\sum_{S_{N}^{(k)}=\pm 1}(\cdots).
\end{equation} 
\subsection{A Bayesian interpretation}
It might be helpful for us to interpret the above 
probability distribution (\ref{eq:gibbs}) 
from the view point of Bayesian inference. 
Actually, one can derive 
the above Boltzmann-Gibbs form $P(\bm{S}^{(k)})$
by means of 
a posterior distribution in Bayesian statistics. 
Let us assume that 
the price change of stock might be caused by $\sigma_{\tau}^{(k)}(t)$ 
due to the microscopic decision making $S_{i}^{(k)}=\pm 1$ for 
trader $i$. 
Then, the following conditional probability 
\begin{equation}
P(\sigma_{\tau}^{(k)}(t) |\bm{S}^{(k)}) \propto 
{\exp}\left(
h_{t}^{(k)}\sum_{i=1}^{N}\sigma_{\tau}^{(k)}(t) S_{i}^{(k)}
\right)
\label{eq:like2}
\end{equation}
denotes a likelihood 
of the result $\sigma_{\tau}^{(k)}(t)$ caused by 
the decision making of traders $\bm{S}^{(k)}$. 
Therefore, in order to 
forecast the decision of 
traders 
$\bm{S}^{(k)}$ for a given 
market data $\sigma_{\tau}^{(k)}(t)$, 
we construct the posterior $P(\bm{S}^{(k)}|\sigma_{\tau}^{(k)}(t))$ by means of 
the Bayesian formula 
\begin{equation}
P(\bm{S}^{(k)}|\sigma_{\tau}^{(k)}(t)) \propto 
P(\sigma_{\tau}^{(k)}(t)|\bm{S}^{(k)})Q(\bm{S}^{(k)}|\{m_{t} ^{(\mu \neq k)}\})
\label{eq:post2}
\end{equation}
where $Q(\bm{S}^{(k)}|\{m_{t} ^{(\mu \neq k)}\})$ stands for the prior distribution. 
For a given set of mean-fields from the other Ising layers 
$\{m_{t}^{(\mu \neq k)}\}$, 
it is naturally accepted to choose the prior as 
\begin{equation}
Q(\bm{S}^{(k)}|\{m_{t} ^{(\mu \neq k)}\}) \propto {\exp}
\left[
\frac{J_{t}^{(k)}}{N_{k}}\sum_{ij}S_{i}^{(k)}S_{j}^{(k)}
+\gamma_{t}^{(k)}\sum_{i}
\left(
\frac{1}{K}\sum_{\mu \neq k}^{K}
c_{k\mu} (t) m_{t}^{(\mu)}
\right)S_{i}^{(k)}
\right]
\label{eq:pri2}
\end{equation}
Then, we have the posterior 
\begin{eqnarray}
\fl && P(\bm{S}^{(k)}|\sigma_{\tau}^{(k)}(t)) \nonumber \\
\mbox{}  & = & 
\frac{{\exp}[
\frac{J_{t}^{(k)}}{N_{k}}\sum_{ij}S_{i}^{(k)}S_{j}^{(k)}
+
h_{t}^{(k)}\sum_{i}\sigma_{\tau}^{(k)}(t)S_{i}^{(k)}
+
\gamma_{t}^{(k)}\sum_{i}
(
\frac{1}{K}\sum_{\mu \neq k}^{K}
c_{k\mu} (t) m_{t}^{(\mu)}
)S_{i}^{(k)}
]
}
{\sum_{\bm{S}^{(k)}}
{\exp}[\frac{J_{t}^{(k)}}{N_{k}}\sum_{ij}S_{i}^{(k)}S_{j}^{(k)}
+
h_{t}^{(k)}\sum_{i}\sigma_{\tau}^{(k)}(t) S_{i}^{(k)}
+
\gamma_{t}^{(k)}\sum_{i}
(
\frac{1}{K}\sum_{\mu \neq k}^{K}
c_{k\mu} (t) m_{t}^{(\mu)}
)S_{i}^{(k)}
]}. \nonumber \\
\label{eq:posterior}
\end{eqnarray}
Thus, 
our model system is described 
by multi-layered Ising model 
in which 
arbitrary two layers (stocks) are coupled through the 
mean-fields.
\subsection{The mean-field equation for instantaneous return}
As it is well-known, 
when ingredients of the system are `fully-connected', 
the partition function $Z$ (the numerator of (\ref{eq:posterior})) is 
evaluated at the saddle point as 
\begin{equation}
Z \simeq 
{\exp}
\left[
N_{k} \Phi (m^{(k)}: J_{t}^{(k)},h_{t}^{(k)},\gamma_{t}^{(k)}, \{m_{t}^{(\mu)}\})
\right]
\end{equation}
in the limit of $N_{k} \to \infty$ with `free energy density' 
\begin{equation}
\Phi =
-\frac{J_{t}^{(k)}}{2}(m^{(k)})^{2} + 
\log 
\cosh 
\left(
J_{t}^{(k)} m^{(k)}
+h_{t}^{(k)} \sigma_{\tau}^{(k)}(t) +  
\frac{\gamma_{t}^{(k)}}{K}\sum_{\mu \neq k}^{K}
c_{k\mu} (t) m_{t}^{(\mu)}
\right). 
\end{equation}
Thus, the saddle point equation $\partial \Phi/\partial m^{(k)}=0$ yields 
\begin{equation}
m^{(k)} = 
\tanh 
\left(
J_{t}^{(k)} m^{(k)}
+h_{t}^{(k)} \sigma_{\tau}^{(k)}(t) +  
\frac{\gamma_{t}^{(k)}}{K}\sum_{\mu \neq k}^{K}
c_{k\mu} (t) m_{t}^{(\mu)}
\right).
\end{equation}
The second and third terms appearing 
in $\tanh(\cdots)$ are 
an external field from 
the market history and 
mean-fields from  
the other Ising layers, respectively. 
For a given non-stationary field $\sigma_{t}^{(k)}(t)$ and 
mean-field $\{m_{t}^{(\mu)}\}$, 
it could not be expected 
that the equilibrium solution for each layer $m^{(k)}$ exists. 
Hence, here we assume that 
such non-equilibrium effects could be built-in 
by dealing with the following time-dependent (naive) mean-field equation for the `instantaneous' returns. 
\begin{equation}
m_{t} ^{(k)} = 
\tanh 
\left(
J_{t-1}^{(k)} m_{t-1}^{(k)}
+h_{t-1}^{(k)} \sigma_{\tau}^{(k)}(t-1) +  
\frac{\gamma_{t-1}^{(k)}}{K}\sum_{\mu \neq k}^{K}
c_{k\mu} (t-1) m_{t-1}^{(\mu)}
\right),\,\, k=1, \cdots, K
\label{eq:update_m}
\end{equation}
By solving the above non-linear maps numerically for a given set of 
past real market history $\{\sigma_{\tau}^{(k)} (t-1)\}$, 
one can forecast the price of each stock 
by means of $\Delta p_{t} =m_{t}^{(k)},\,\,k=1,\cdots,K$ 
simultaneously. 
\subsection{Hyper-parameter estimation from non-stationary time-series}
In order to use the mean-field equation (\ref{eq:update_m}) for forecasting the stock prices,  systematic estimation for the so-called 
hyper-parameters $(J_{t}^{(k)},h_{t}^{(k)},\gamma_{t}^{(k)})$ appearing in the right hand side of 
(\ref{eq:update_m}) is needed. 
In the literature of 
probabilistic information processing, 
say, in Bayesian image restoration \cite{Inoue}, 
one cannot use the mean-square error as 
a cost function because 
it needs the `true (original) image' to be constructed. 
Therefore,  
we usually use the 
marginal likelihood 
as the cost function 
to estimate the hyper-parameters.  
However, fortunately in our present model system, 
one can utilize the past market history, let us to say, 
the `true returns' in the past 
$\Delta q_{l}^{(k)} \equiv q_{l}^{(k)}-q_{l-1}^{(k)},\,\,
l=1,\cdots,t-1$, and 
it means that the `cumulative error' could be defined as  a 
cumulative mean-square error between the true observable and the estimate by 
\begin{eqnarray}
\mathcal{E}_{k}  & \equiv &    
 \frac{1}{2}
\sum_{l=1}^{t-1}
\left(
\overline{\Delta q_{l}^{(k)}} -  
\overline{\Delta p_{l}^{(k)}}
\right)^{2} \nonumber \\
\mbox{} & = &   
\frac{1}{2}
\sum_{l=1}^{t-1}
{\Biggr [}
\overline{\Delta q_{l}^{(k)}} -   
\tanh 
{\Biggr (} 
J_{t-1}^{(k)} \overline{\Delta q_{l-1}^{(k)}}  +   h_{t-1}^{(k)} \sigma_{\tau}^{(k)}(t-1) +    
\frac{\gamma_{t-1}^{(k)}}{K}\sum_{\mu=1}^{K}
c_{k\mu} (t-1) \overline{\Delta q_{l-1}^{(\mu)}} 
{\Biggr )}
{\Biggr ]}^{2}
\label{eq:cost_to_parameters} 
\end{eqnarray}
where we defined the `forecasted return' $\Delta p_{l}^{(k)} \equiv p_{l}^{(k)}-p_{l-1}^{(k)},\,\,
l=1,\cdots,t-1$ and the time-averages 
\begin{equation}
\overline{\Delta q_{l}^{(k)}} \equiv \frac{1}{M}
\sum_{i=l-M +1}^{l}  (q_{i+1}^{(k)}-q_{i}^{(k)}),\,\,
\overline{\Delta p_{l}^{(k)}} \equiv \frac{1}{M}
\sum_{i=l-M +1}^{l}  (p_{i+1}^{(k)}-p_{i}^{(k)})
\label{eq:qp}
\end{equation}
for the time window with width $M$. 
To obtain the last line in the above 
equation (\ref{eq:cost_to_parameters}), we used 
\begin{equation}
\overline{\Delta p_{l}^{(k)}} \simeq 
m_{l}^{(k)}=
\tanh 
\left(
J_{t-1}^{(k)} m_{l-1}^{(k)}
+h_{t-1}^{(k)} \sigma_{\tau}^{(k)}(t-1) +  
\frac{\gamma_{t-1}^{(k)}}{K}\sum_{\mu \neq k}^{K}
c_{k\mu} (t-1) m_{l-1}^{(\mu)}
\right)
\end{equation}
and replaced the $m_{l-1}^{(k)},\, k=1,\cdots,K$  
appearing in $\tanh(\cdots)$ by the corresponding observables 
$\overline{\Delta q_{l-1}^{(k)}},\,k=1,\cdots,K$ because 
one can actually use these values before forecasting.

Hence, we should infer these hyper-parameters $(J_{t}^{(k)},h_{t}^{(k)},\gamma_{t}^{(k)})$ 
from the past data set in the financial market by the gradient descent learning 
\begin{equation}
J_{t}^{(k)} =
J_{t-1}^{(k)} - \eta\, \frac{\partial \mathcal{E}_{k}}{\partial J_{t-1}^{(k)}},\,\,
h_{t}^{(k)} = 
h_{t-1}^{(k)} - \eta\, \frac{\partial \mathcal{E}_{k}}{\partial h_{t-1}^{(k)}},\,\,
\gamma_{t}^{(k)} = 
\gamma_{t-1}^{(k)} - \eta\, \frac{\partial \mathcal{E}_{k}}{\partial \gamma_{t-1}^{(k)}}, 
\label{eq:gammat}
\end{equation}
where 
$\eta$ is a learning rate. 
In the next section, we examine 
the above forecasting framework for empirical (intra-day) data sets in which a crisis appears. 
\section{Performance evaluation of forecasting for empirical high-frequency data sets}
\label{sec:se5}
\begin{figure}[ht]
\begin{center}
\includegraphics[width=8cm]{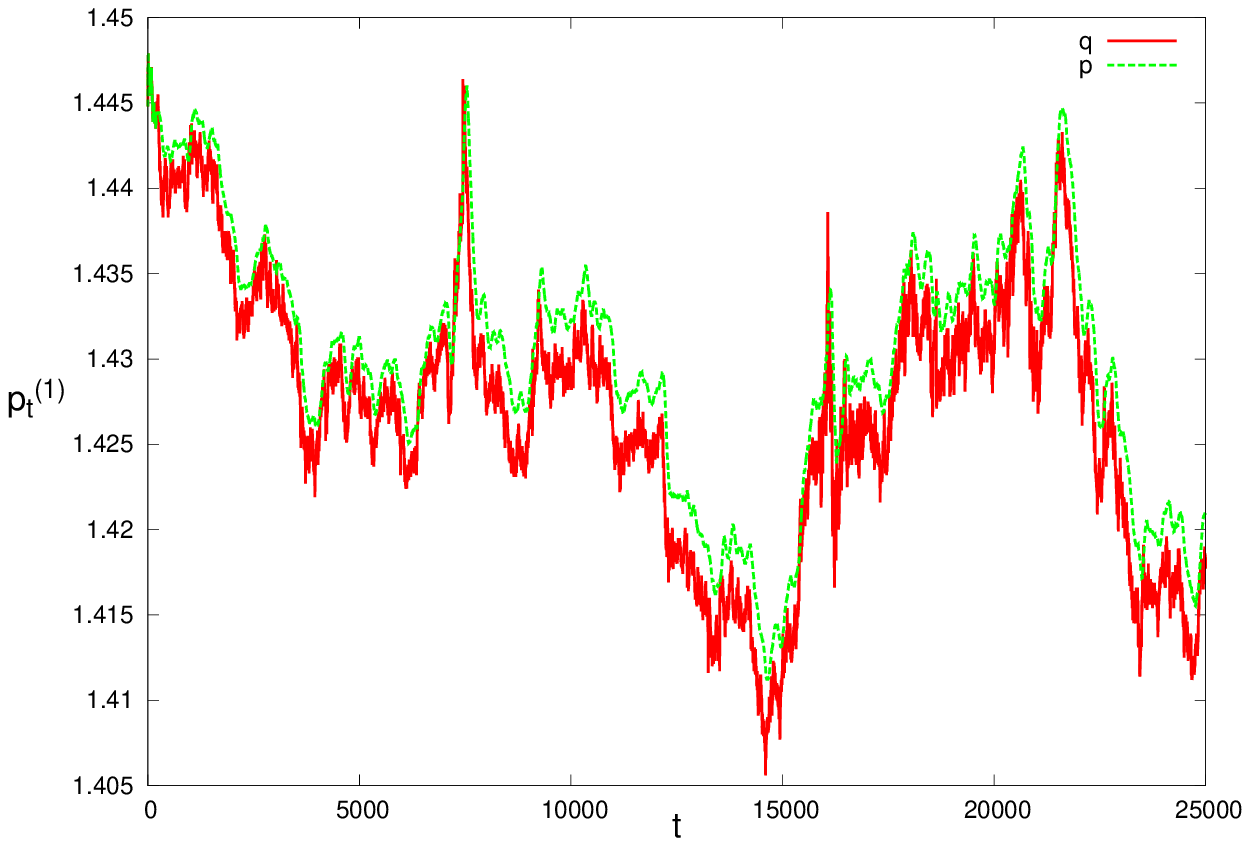} \hspace{-0.5cm}
\includegraphics[width=8cm]{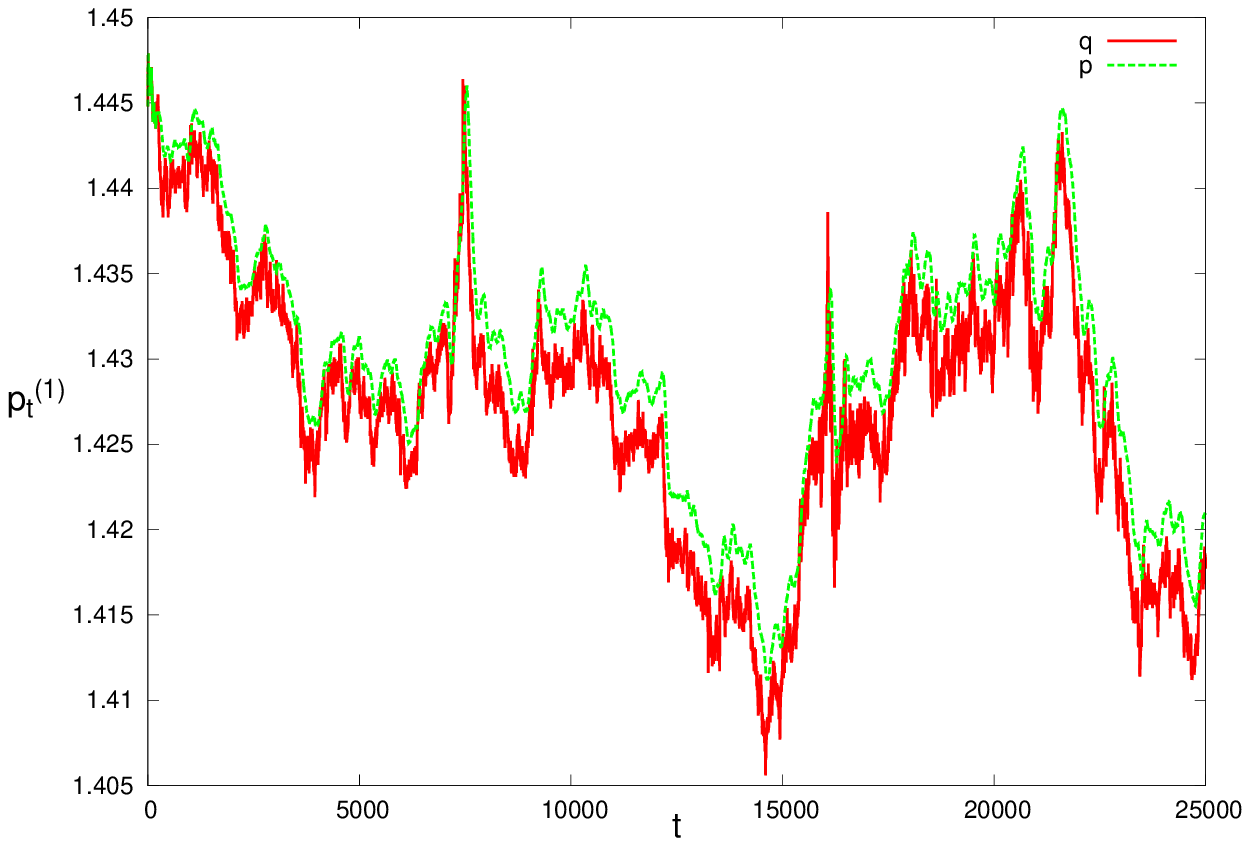} \\
\includegraphics[width=8cm]{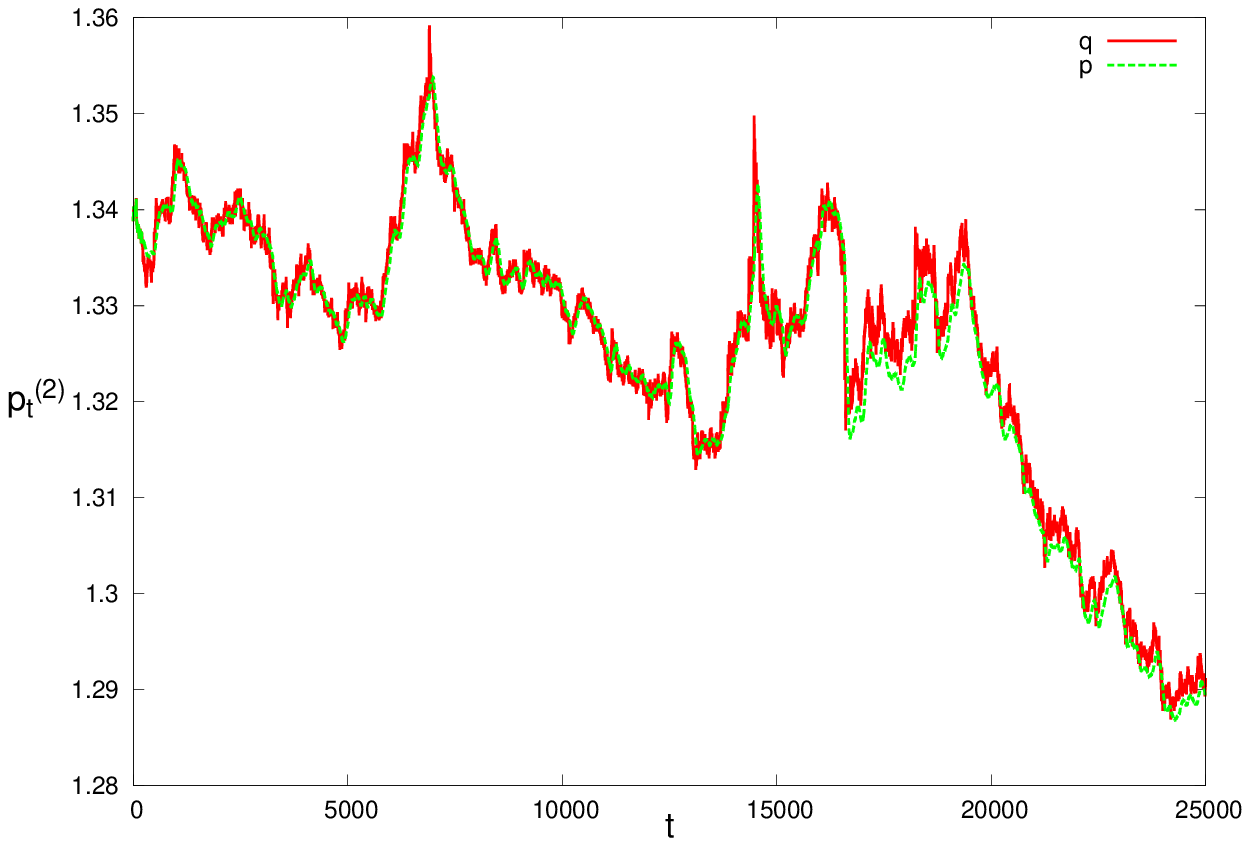} \hspace{-0.5cm}
\includegraphics[width=8cm]{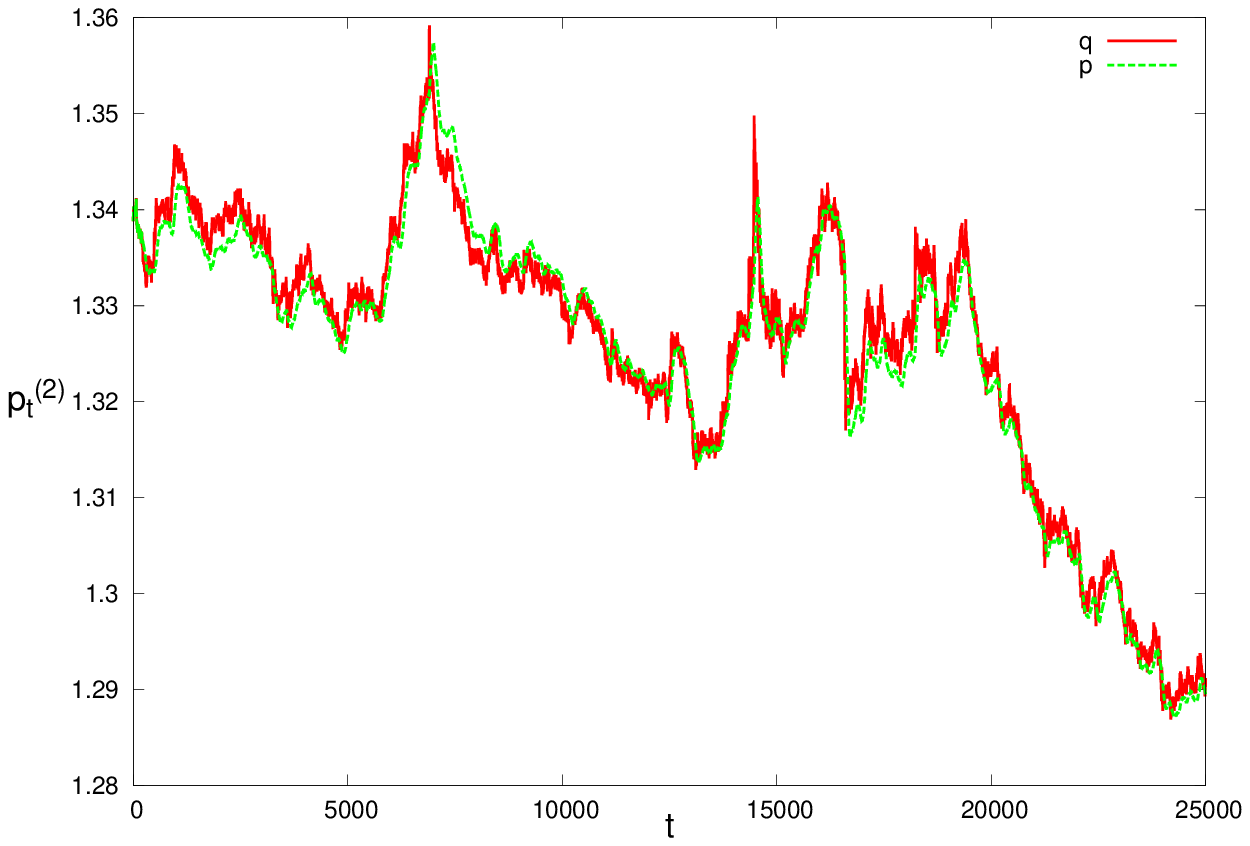} \\
\includegraphics[width=8cm]{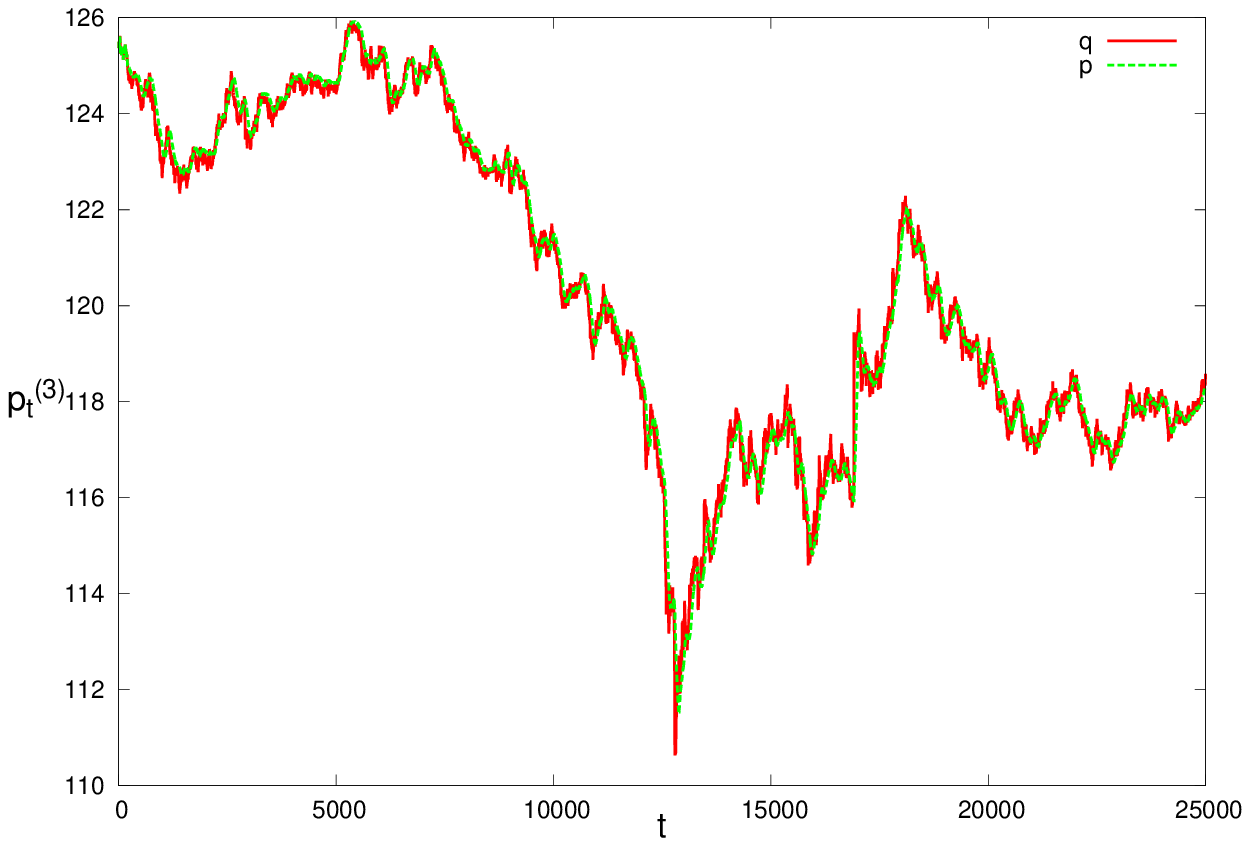} \hspace{-0.5cm}
\includegraphics[width=8cm]{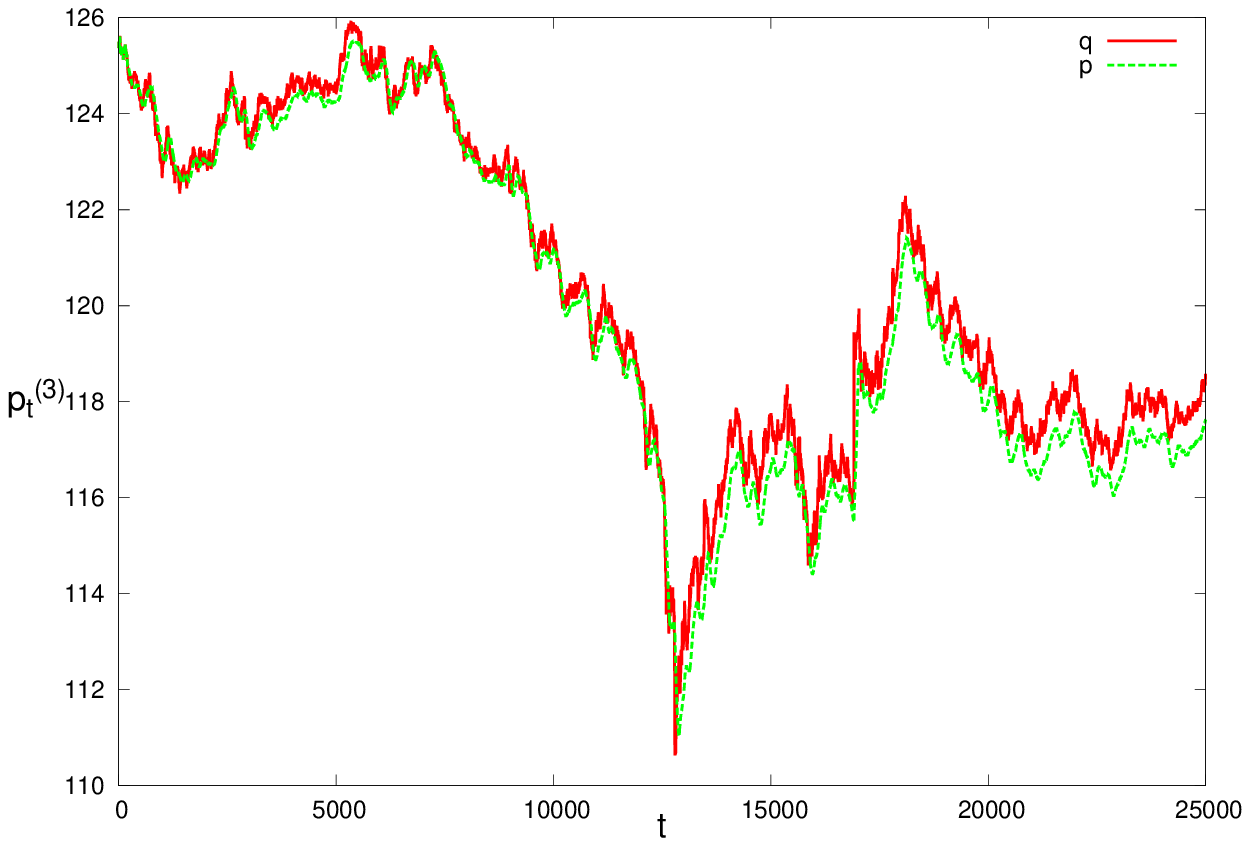}
\end{center}
\caption{\footnotesize 
From the top to the bottom, EUR/AUD ($k=1$),
real value $q_{t}^{(k)}$ and 
its prediction $p_{t}^{(k)}$ are plotted 
for EUR/CAD ($k=2$), 
EUR/JPY ($k=3$) exchange rates 
from 27th April 2010 to 13th May 2010. 
The panels in the left column are obtained 
by `stock-wise optimization', whereas the panels in the right 
are calculated by `pair-wise optimization' for 
the correlation strengths. 
We set the width of time window and the learning rate as 
$M=\tau=100, \eta=0.01$. (COLOR ONLINE)
}
\label{fig:fg21}
\end{figure}

In this section, we check the usefulness of our prediction model 
with cross-correlation in stocks. 
As a simple examination, 
for the case of three different time-series $K=3$, 
we check the accuracy of 
our prediction procedure.  
In section \ref{sec:se2},  
we visualized 
200 stocks which 
are given as daily data during a crisis. 
However, the number of data points is not enough 
for our forecasting procedure. 
Hence, here we pick up EUR/AUD ($k=1$),
EUR/CAD ($k=2$), 
EUR/JPY ($k=3$) exchange rates (EUR: Euro, CAD: Canadian dollar, 
AUD: Australian dollar, JPY: Japanese yen), which are given as high-frequency tick-by-tick data, 
from 27th April 2010 to 13th May 2010 as real values $q_{t}^{(k)}$.  
We plot those three true time series in Figure \ref{fig:fg21} as solid lines.  
We observe that these time series posses a crisis which 
corresponds to Greek crisis in spring 2010. 
In the left panels of Figure \ref{fig:fg21}, 
the resulting prices $p_{t}^{(k)}$ predicted by 
our model are shown. 
We set 
the time window size as $M=\tau=100$ and the learning rate as $\eta=0.01$. 
Of course, we can choose the leaning rate as `adaptive one' like $\eta = \eta (t)$, however, in this paper we set the value to a positive constant. 
From these panels, 
we find that 
our prediction procedure works well 
and the error is only a few percent of the 
average value of the rate. 
\mbox{}
\subsection{Pair-wise optimization for hyper-parameters of correlation strengths}
We next examine the following slight modification 
\begin{equation}
\frac{\gamma_{t}^{(k)}}{K}
\sum_{\mu \neq k}^{K}
c_{k\mu} (t-1) \overline{\Delta q_{l-1}^{(\mu)}} \to 
\frac{1}{K}
\sum_{\mu \neq k}^{K}
\gamma_{t}^{(k\mu)} c_{k\mu} (t-1) \overline{\Delta q_{l-1}^{(\mu)}}
\label{eq:modification}
\end{equation}
in (\ref{eq:update_m})-(\ref{eq:cost_to_parameters}). 
Namely, 
the strength of correlation between Ising layers 
is evaluated independently for each pair of layers. 
Due to this modification, the learning equation for $\gamma_{t}^{(k)}$ in (\ref{eq:gammat}) should be corrected as 
\begin{equation}
\gamma_{t}^{(k\mu)} =
\gamma_{t-1}^{(k\mu)} -\eta\, 
\frac{\partial \mathcal{E}_{k}}{\partial \gamma_{t-1}^{(k\mu)}},\, \mu \neq k. 
\label{eq:mod_gamma}
\end{equation}
From now on, the forecasting model with the above modification (\ref{eq:modification})(\ref{eq:mod_gamma}) is 
referred to as `pair-wise optimization', whereas 
the original version (\ref{eq:update_m})(\ref{eq:cost_to_parameters}) is 
called as `stock-wise optimization' for hyper-parameters for the correlation strengths. 

We plot the result in the right panels in Figure \ref{fig:fg21}. 
From these panels, 
we are confirmed that the pair-wise optimization makes the result worse against our expectation. 
Especially, for large $t$ regime, 
the gap between the stock-wise optimization and the pair-wise optimization becomes large. 
From the result, we should consider the optimal number of hyper-parameters 
to fit the model to the time-series without over-fitting. 
\subsection{Flows of hyper-parameters during a crisis}
\begin{figure}[ht]
\begin{center}
\includegraphics[width=8cm]{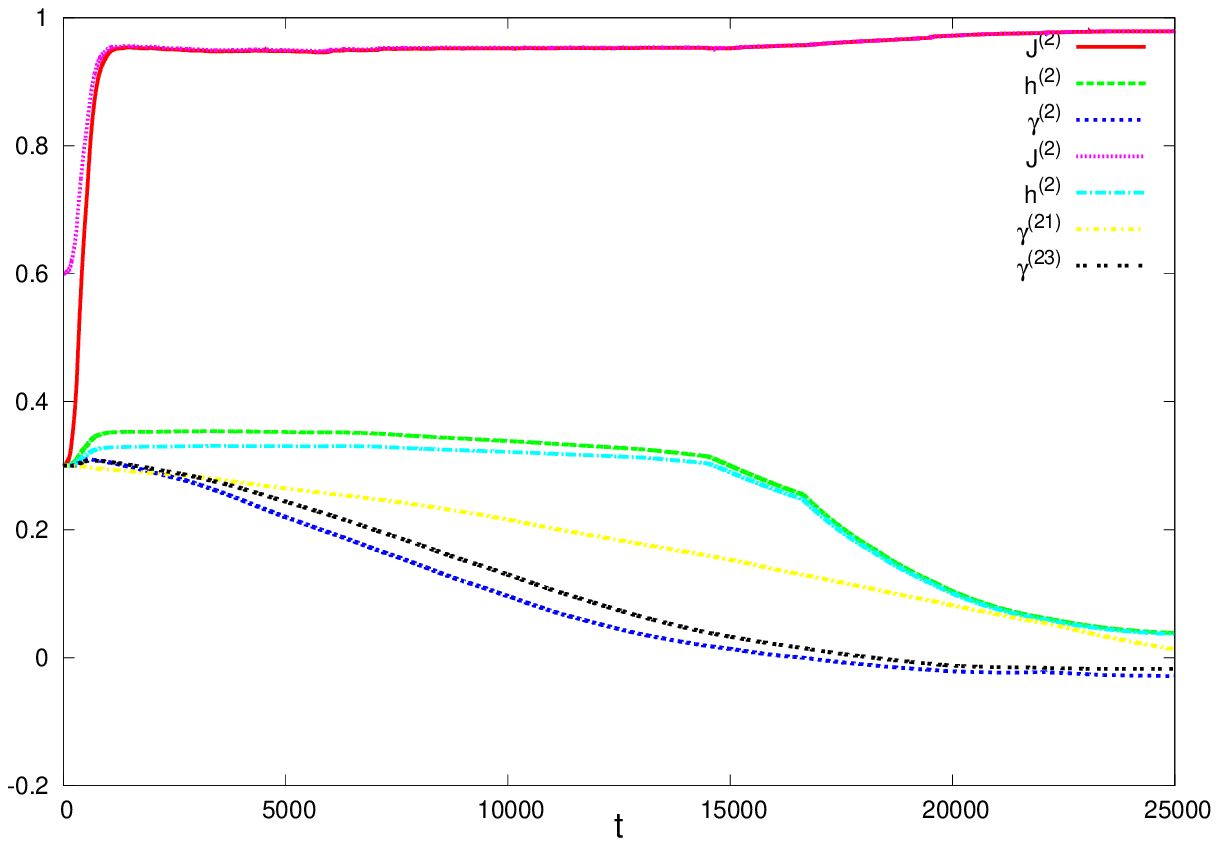} \hspace{-0.5cm}
\includegraphics[width=8cm]{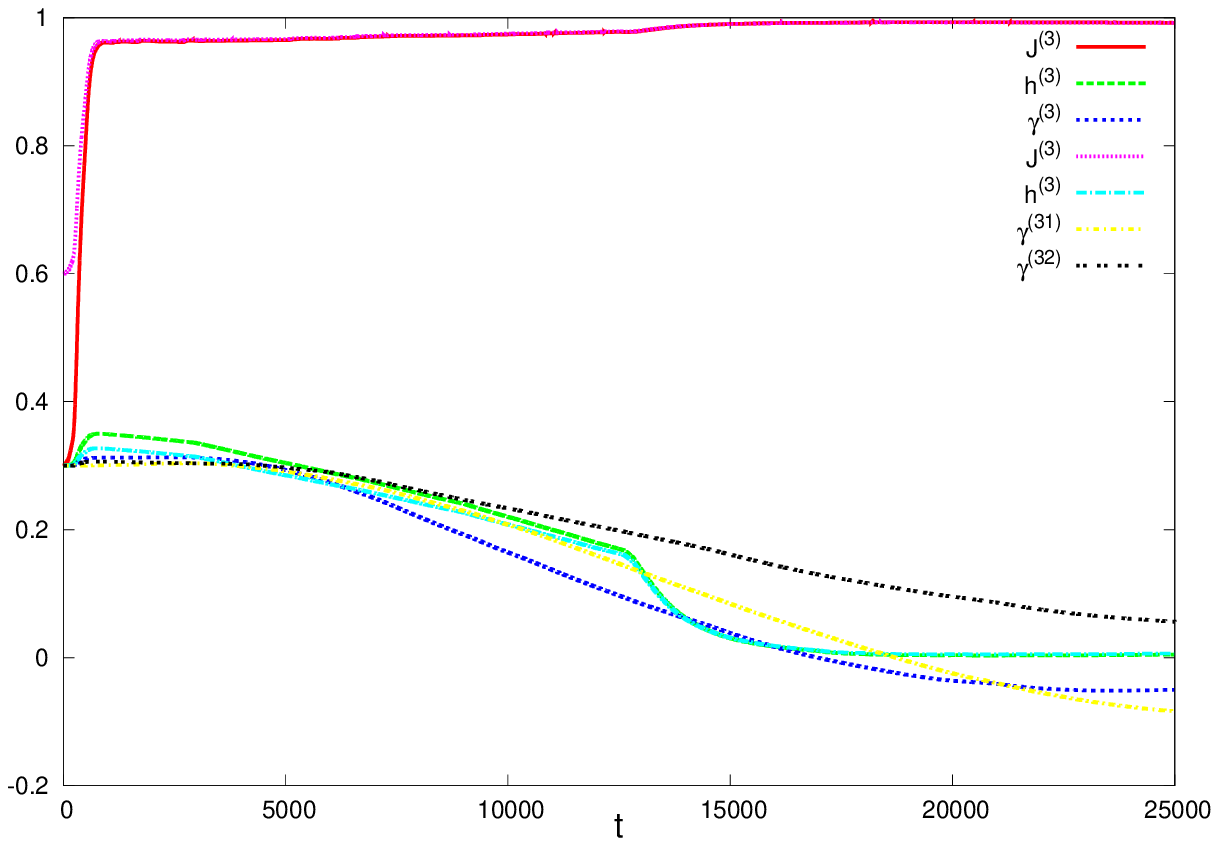}
\end{center}
\caption{\footnotesize 
Dynamics of macroscopic parameters 
$(J_{t}^{(k)},h_{t}^{(k)},\gamma_{t}^{(k)})$ for `stock-wise optimization' 
(The top three line captions) 
and $(J_{t}^{(k)},h_{t}^{(k)},\gamma_{t}^{(k\mu)})$ for `pair-wise optimization' (The bottom four line captions). 
We show the result for $k=2$ (left panel) and $k=3$ (right). (COLOR ONLINE)}
\label{fig:fg22}
\end{figure}

Finally, we consider the dynamical behaviour 
of hyper-parameters during the crisis. 
We show the results in Figure \ref{fig:fg22}. 
These panel show the dynamics of hyper-parameters 
$(J_{t}^{(k)},h_{t}^{(k)},\gamma_{t}^{(k)})$ for 
the stock-wise optimization and 
and $(J_{t}^{(k)},h_{t}^{(k)},\gamma_{t}^{(k\mu)})$ for the pair-wise optimization. 

From this panel, 
we clearly find that for both cases $(k=2,3)$ of exchange rates in our forecasting model described by stock-wise optimization,  
the hyper-parameters $J^{(k)},h^{(k)}$ converge to 
$h^{(k)} =0$ and $J^{(k)}=1$, respectively. Then, the $\gamma_{t}^{(k)}$ 
converges to zero for $k=2$, which corresponds to the critical point of order-disorder phase transition in the infinite-range ferromagnetic Ising model. 
However, 
for $k=3$ it converges to slightly negative value across the zero. 
Hence for $k=3$ the stock-wise optimization 
does not detect the critical behaviour,  and the deviation from the critical point becomes large as $t$ goes on. 
On  the other hand, the model with the pair-wise optimization 
also detects the critical behaviour for $k=2$, however, 
the behavior of the hyper-parameters does not reflect the critical point (crisis) for $k=3$. 

We can partially figure out these results in 
the original time series shown in Figure \ref{fig:fg21}. 
From this figure, we can observe that 
the exchange rate $k=3$ was recovered from the crisis relatively faster than $k=2$, whereas 
it was still in a crisis for the exchange rate $k=2$ even in large $t$ regime. 
Actually, in this period, 
Japanese yen was strong, and at the crisis, 
this tendency of strong yen was enhanced. 
However, the term of extreme strong yen is not so long and 
the value of Euro against yen was recovered quickly. 
On the other hand, 
continuous drops of Euro 
against Canadian dollar was serious 
after the crisis and it could not escape from a `crush domain' even at $t=250000$ (ticks). 
In this sense, 
the deviation from the critical point should be also observed through our model 
and actually it was confirmed in Figure \ref{fig:fg22}. 
\section{Discussions and concluding remarks}
There are many ways to visualize the co-movements of stocks, and using MDS is one of them. When the MDS studies
were performed with daily data \cite{Tilak}, we found that it was easier to visualize or detect
specific sectors, strongly correlated pairs and market events. It was suggested that this
type of plots using daily data may be used in designing strategies of “pairs trade”  or identifying clusters or detecting market trends. It was also shown in Ref. \cite{Tilak}, that we could follow the evolution of the market and/or trace the movements of particular companies (e.g., Lehmann Brothers) with respect to the rest of the market, before or after a market crisis. 

In this paper, in order to show and forecast some cascade in financial systems, 
we visualized the correlation of each pair of stocks in two-dimension using MDS. 
We also proposed a theoretical framework based on the multi-layered Ising model to predict several time-series 
simultaneously by using cross-correlations in financial markets. 
Actually,  in this paper, we showed that the knowledge of statistical mechanics of information (see {\it e.g.} \cite{Nishimori}) could be applicable to 
the research topics outside of information processing, namely, 
quantitative finance or economics. 

Finally, we would like to mention 
several remarks concerning our future direction. 
\subsection{On the time window size}
In our model system, 
we set $M=\tau=100$ for the width of time window to 
evaluate several statistics in our forecasting model (see (\ref{eq:trend}) and (\ref{eq:qp})). 
However, we should chose these lengths more carefully. 
Recently, Livan, Inoue and Scalas \cite{LIS} 
examined the effect of non-stationarity of time series on the portfolio 
optimization by using several statistical test including some knowledge of the random matrix theory \cite{Giacomo}, 
and they found that the longer time window 
does not always give the better estimate of the true value of the portfolio.  
This implies that there might exist some optimal size of 
window to construct the forecasting model.  
\subsection{The turnover}
\begin{figure}[ht]
\begin{center}
\includegraphics[width=8.1cm]{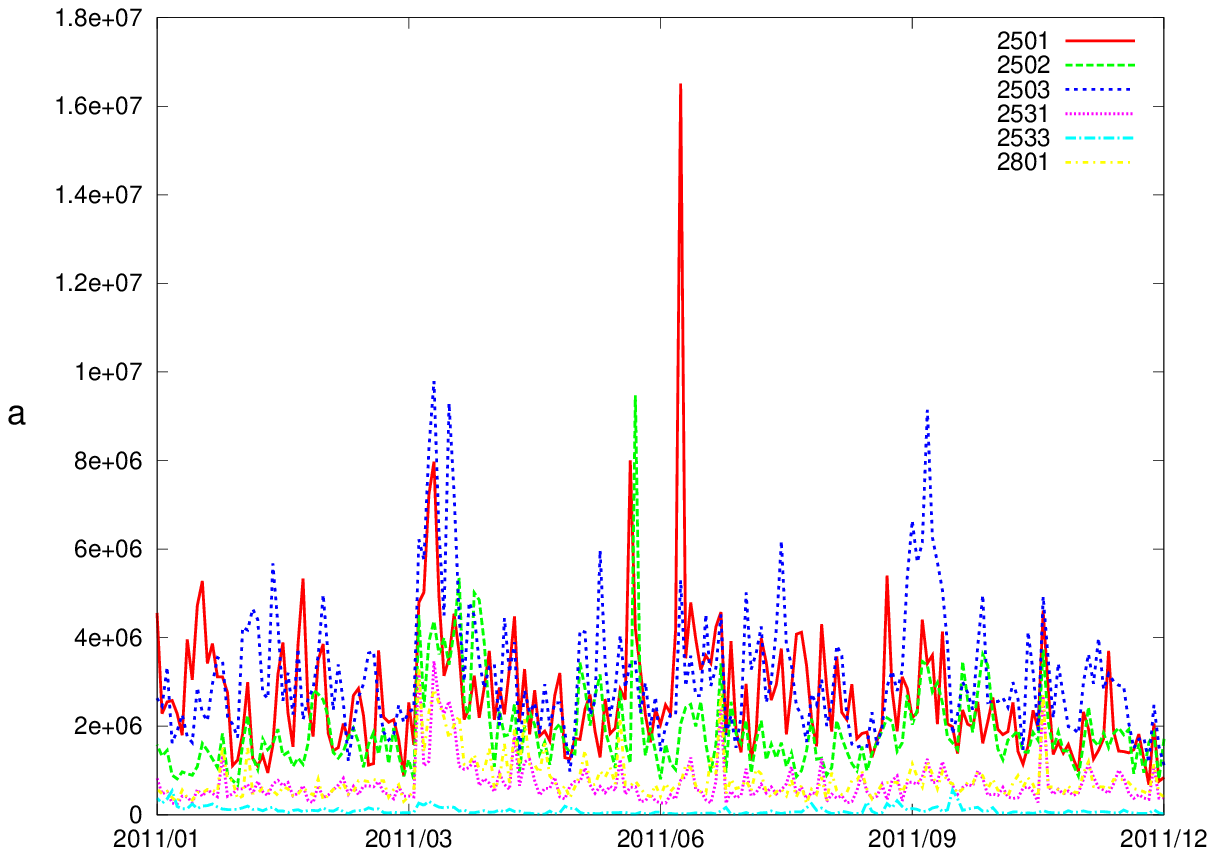} \hspace{-0.5cm}
\includegraphics[width=8.1cm]{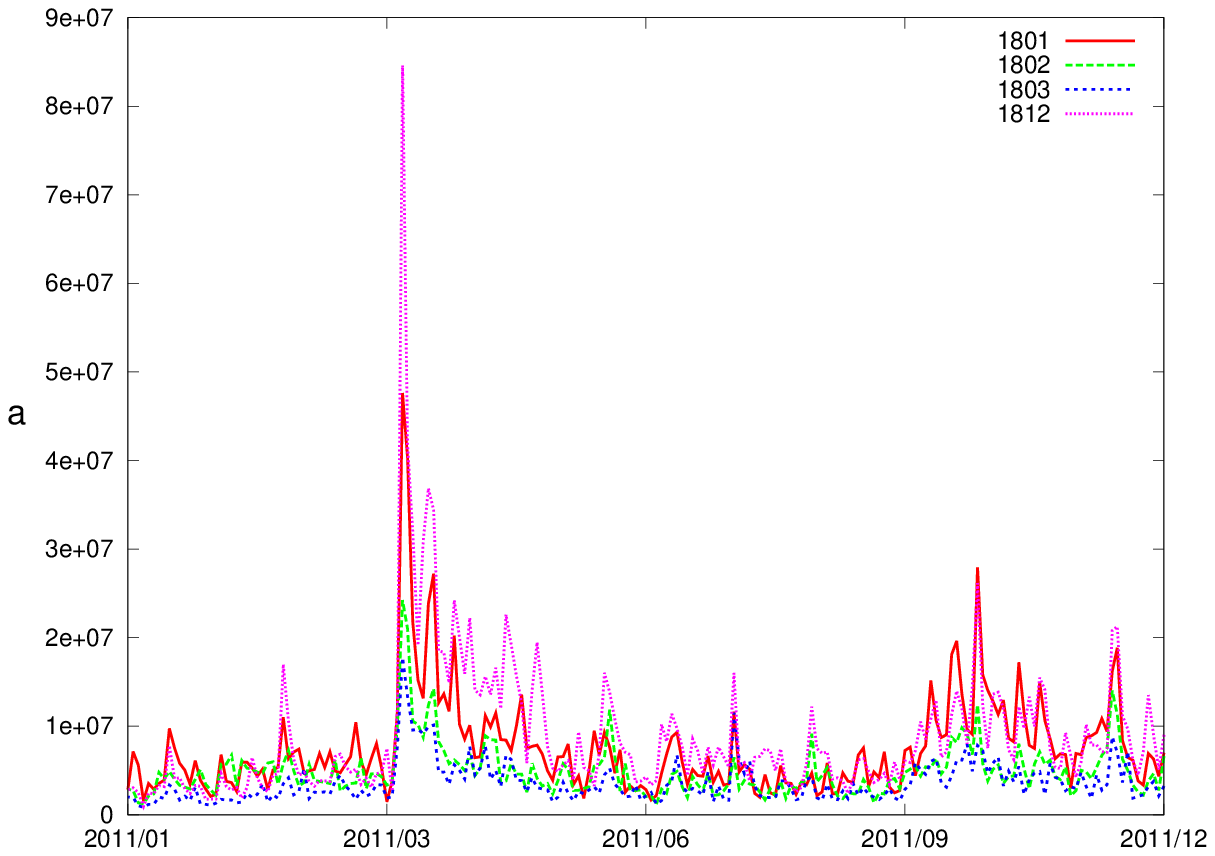}
\end{center}
\caption{\footnotesize 
The turnover as a function of time (day). 
The number of caption denotes the ID for each company: 
({\it i.e.} {\tt 2501}: Sapporo Breweries, 
{\tt 2502}: Asahi Breweries, {\tt 2503}: Kirin Holdings, 
{\tt 2531}: Takara Holdings, {\tt 2533}: Oenon Holdings, 
{\tt 2801}: Kikkoman Corporation, 
{\tt 1801}: Taisei Corporation, 
{\tt 1802}: Obayashi Corporation, 
{\tt 1803}: Shimizu Corporation, 
{\tt 1812}: Kajima Corporation. These IDs can be checked at the web site \cite{Yahoo}) (COLOR ONLINE)}
\label{fig:fg9}
\end{figure}

The turnover, 
namely, 
the total number of volume being dealt with for a stock 
at time $t$: 
\begin{equation}
a_{k} (t) \equiv 
\psi_{+}(k,t)+\psi_{-}(k,t) =\sum_{i=1}^{N_{k}}v_{it}^{(k)}
\end{equation}
might be a good indicator for the financial crisis (see also (\ref{eq:psi})). 
Actually, in Figure \ref{fig:fg9}, 
we show the empirical plot of the turnover 
for several stocks during the crisis in Japan 2011. 
From this figure, 
we find that the turnover takes a sharp peak 
around the shock (11th March 2011), which was remarkably 
observed in stocks of construction industries (see the right panel of Figure \ref{fig:fg9}). 
In our preliminary study  \cite{Murota}, 
we made a model to estimate the turnover 
by means of three state Ising model in which 
each spin can take zero (`staying') besides $\pm 1$ for `selling' or `buying'. 
\subsection{Suitable estimator for non-synchronous time series}
The data sets used in the forecasting examination 
are not daily data but high-frequency data. 
Those are `non-synchronous' time series, 
namely, 
there is no one-to-one correspondence in 
time axis in arbitrary two stocks. 
In this paper, for 
evaluating, say, 
$\Delta r_{i}(t) \Delta r_{j}(t^{'})$ where 
$t\neq t^{'}$, 
we chose the $t^{'}$ simply by 
\begin{equation}
t^{'} = 
{\rm argmin}_{t} \| t-t^{'} \|
\end{equation}
and of course, it might posses some evaluation error. 
In this sense (namely, `strict sense'), 
the Pearson estimator is not suitable to 
evaluate the cross-correlation, 
and we should use another way, say, 
the so-called {\it Hayashi-Yoshida estimator} \cite{Hayashi,Tilak}. 
\subsection{The inverse-Ising problem}
In our forecasting model, 
we assumed that the traders are fully-connected. 
However, the graph topology is important for us to 
consider the communities in the markets. 
Hence, the procedure to estimate the adjacency matrix 
from the empirical data (behavior of traders) should be done. 
Definitely,  it is formulated as an `inverse Ising problem'. 

The studies related to the above four issues 
are now on-going and 
we will report the results at the meeting  
if we obtain the preliminary. 
\section*{Acknowledgment}
The authors would like to thank 
Enrico Scalas, Giacomo Livan, Fr$\acute{\rm e}$d$\acute{\rm e}$ric Abergel for 
fruitful discussion and useful comments. 
SS and JI thank Saha Institute of Nuclear Physics 
for their support during our stay in Kolkata. 
One of the authors (JI) thanks 
Basque Center for Applied Mathematics,  
\'{E}cole Centrale Paris for their warm hospitality. 
JI also acknowledges the financial support by 
Grant-in-Aid for Scientific Research (C) 
of Japan Society for 
the Promotion of Science, No. 22500195 (2010-2012) and 
No.25330278 (2013-2015). AC is grateful to Hokkaido University for support during his stay in Sapporo.

\end{document}